\documentclass[letterpaper]{article}
\usepackage{spconf,graphicx,tabularx,booktabs,amsmath,balance,enumitem,makecell}
\usepackage[square,sort&compress,comma,numbers]{natbib}
\usepackage{xcolor}
\usepackage[normalem]{ulem}
\usepackage{multirow}
\usepackage{tabularx}

\setlist[itemize]{leftmargin=*}

\makeatletter

\title{M\MakeLowercase{u}SE-ing on the impact of utterance ordering on crowdsourced emotion annotations}
\def\@name{ Mimansa Jaiswal$^\ast$, Zakaria Aldeneh$^\ast$, Cristian-Paul Bara$^\ast$, Yuanhang Luo$^\ast$, Mihai Burzo$^\dagger$, \\
\emph{Rada Mihalcea$^\ast$, Emily Mower Provost$^\ast$}}
\makeatother
\address{ $^\ast$University of Michigan--Ann Arbor \\ $^\dagger$University of Michigan--Flint}

\begin{document}
\ninept
\maketitle

\begin{abstract}
Emotion recognition algorithms rely on data annotated with high quality labels. However, emotion expression and perception are inherently subjective.  There is generally not a single annotation that can be unambiguously declared ``correct.''  As a result, annotations are colored by the manner in which they were collected.  In this paper, we conduct crowdsourcing experiments to investigate this impact on both the annotations themselves and on the performance of these algorithms.  We focus on one critical question: the effect of context.  We present a new emotion dataset, Multimodal Stressed Emotion (MuSE), and annotate the dataset using two conditions: randomized, in which annotators are presented with clips in random order, and contextualized, in which annotators are presented with clips in order.
We find that contextual labeling schemes result in annotations that are more similar to a speaker's own self-reported labels and that labels generated from randomized schemes are most easily predictable by automated systems.
\end{abstract}



\section{Introduction}
\vspace{-2mm}
Emotion technologies, both recognition and synthesis, are heavily dependent on having reliably annotated emotional data, annotations that describe the observed emotional display.  The hope is often that these annotations capture the speaker's true underlying state.  Yet, in practice, this true \emph{felt sense} emotion is unknown, and researchers must resort to manual labeling of data.  The hope is that these manual labels are sufficiently ``correct'' to enable the training and evaluation of emotion technologies.  One method of ensuring quality labels has been to require the participation of expert raters. However, it can be both expensive and time consuming to hire expert raters. More recently, researchers have embraced crowdsourcing services (e.g., \textit{Amazon Mechanical Turk}) to efficiently collect annotations from non-expert workers in a cost-effective and timely manner \cite{soleymani2010crowdsourcing}. Once collected, annotations from non-expert workers are aggregated to form ground-truth labels that are used for training and evaluating automated systems. However, the method through which these annotations are collected can profoundly impact the behavior of the annotators.  In this paper, we study how the setup of a crowdsourcing task can influence both the collected emotion labels as well as the performance of classifiers trained using these labels.

The effective use of crowdsourcing for collecting reliable emotion labels has been an active research topic. Burmania et al. investigated the trade-off between the number of annotators and underlying reliability of the annotations~\cite{burmania2016tradeoff}. 
Other work has looked at quality-control techniques to improve the reliability of annotations.  For example, Soleymani et al. used qualification tests to filter out spammers and retain high-quality annotators~\cite{soleymani2010crowdsourcing}. Burmania et al. investigated the use of gold-standard samples to monitor annotators' reliability and fatigue~\cite{burmania2016increasing}.

However, variability also results from context, relevant past information that provides cues as to how to interpret an emotional display. Context, such as tone, words, expressions can affect how individuals perceive emotion \cite{laplante2003things}.
Context is also implicitly included in the labeling schemes of many of the most common emotion datasets (e.g., IEMOCAP~\cite{busso2008iemocap} and MSP-Improv~\cite{busso2017msp}) because annotators rate each utterance (or time period) in order.  That means that annotators are influenced by information that they recently observed~\cite{yannakakis2017ordinal}.  However, emotion recognition systems are often trained over single utterances~\cite{aldeneh2017using,abdelwahab2017incremental,mirsamadi2017automatic, sarma2018emotion}, leading to a mismatch in the information available to annotators and to classification systems. 

In this work, we study the difference between annotations obtained for audio clips when emotional displays are presented to annotators with context and when presented randomly. In both cases, annotators are affected by the emotion displays that they have recently observed \cite{qiao2017transient, russell2017emotion}.  However, only in the contextual presentation there is also a cohesive story. 
We investigate the following research questions:

\begin{itemize}
\itemsep0em
\item Q1: Is there a significant difference between annotations obtained from random and contextual presentations?

\item Q2: Are annotations obtained from contextual presentations more similar to a speaker's own self-reported labels than those from random presentations?

\item Q3: Is there a significant difference between the inter-rater agreements obtained from random and contextual presentations?

\item Q4: How does the performance of an emotion recognition system, operating on single utterances, vary given annotations obtained from random and contextual presentations?

\item Q5: How does the performance gain of an emotion recognition system operating across multiple utterances vary given different amounts of context (defined as number of prior utterances) and labels obtained from random and contextual presentations?

\end{itemize}

This paper is organized as follows. First, we introduce the dataset and explain the design, collection and post-processing procedures. Then, we present an analysis of the dataset and the collected corpus labels. We then present the results of a state-of-the-art speech-based emotion classification system trained on the random presentation vs. the contextual presentation labels.
The findings from this work will provide insight into performance implications of emotion recognition system given mismatches between the amount of context provided to the annotators generating the labels and the ultimate classification system.\looseness=-1

\vspace{-3mm}
\section{Dataset}
\vspace{-2mm}
\label{section:Dataset}
\subsection{Data Collection}
\vspace{-1mm}
We introduce the Multimodal Stressed Emotion (MuSE) dataset, designed to understand how stress and emotion interplay in spoken communication. The dataset consists of fifty-five recordings from twenty-eight participants, each recorded under two conditions, stressed and not-stressed (one subject participated in only the stressed condition).  The stress condition was recorded during the final exam period at the University of Michigan, the not-stressed condition was recorded after exams concluded.  The emotion component was generated through video stimuli, sampled from \cite{gross1995emotion} and \cite{schaefer2010assessing} and through emotionally evocative monologue topics \cite{aron1997experimental}.  The data used in this study are a subset of the corpus. Table \ref{dataset_detail} shows the questions used to evoke emotions, which fall under following sections: (a) icebreaker; (b) non-neutral (c) non-neutral; (d) non-neutral; (e) ending. The non-neutral sections (b), (c) and (d) were presented in random order using prompts from each of the categories: positive, negative, and intensity.  In each case, one question was used in the stress recording and the other was used in the non-stress recording. Each participant was asked to rate his/her emotions after the completion of each section using the scales of activation (calm vs. excited) and valence (positive vs. negative).  We refer to these annotations as the \emph{self-report annotations}.

\vspace{-0.8em}

\subsection{Data Preprocessing}
\vspace{-0.7em}
The monologues in each section are divided into utterances. However, since the monologues are spontaneous, often there is not a clear sentence boundary. We create utterances by identifying prosodic or linguistic boundaries in spontaneous speech as defined by~\cite{kolavr2008automatic}: (a) a clear sentence boundary (full stop or exclamation); (b) a change of context after filler words, or revision of sentence; (c) an extended pause (i.e., a silence greater than three seconds); or (d) filler or example words instead of a full stop.

The dataset contains 2,648 utterances with a mean duration of 12.44 $\pm$ 6.72 seconds (Table~\ref{data_stat}). The mean length of stressed utterances ($11.73\pm5.77$ seconds) is significantly different from that of the non-stressed utterances ($13.30\pm6.73$ seconds).  

We perform data selection, excluding utterances that are shorter than $3$-seconds and longer than $35$-seconds ($2.8$\% of the original data). This is because short segments may not have enough information to capture emotion, and longer segments can have variable emotion.  This results in 2,574 utterances.
\vspace{-0.7em}

\begin{table}[t]

\caption{Emotion elicitation questions.}
\label{dataset_detail}
\vspace{1mm}
\footnotesize
\begin{tabular}{p{8.15cm}}
\toprule

\centerline{\textbf{Icebreaker}}
\vspace{-3pt}
\begin{enumerate}
\itemsep0em 
\item Given the choice of anyone in the world, whom would you want as a dinner guest?
\item Would you like to be famous? In what way?
\end{enumerate} \\
\hline
\\
\centerline{\textbf{Positive}}
\vspace{-3pt}
\begin{enumerate}
\itemsep0em 
\item For what in your life do you feel most grateful?
\item What is the greatest accomplishment of your life? 
\end{enumerate} \\
\hline
\\
\centerline{\textbf{Negative}}
\vspace{-3pt}
\begin{enumerate}
\itemsep0em 
\item If you could change anything about the way you were raised, what would it be?
\item Share an embarrassing moment in your life.
\end{enumerate} \\
\hline
\\
\centerline{\textbf{Intensity}}
\vspace{-3pt}
\begin{enumerate}
\itemsep0em 
\item If you were to die this evening with no opportunity to communicate with anyone, what would you most regret not having told someone?
\item Your house, containing everything you own, catches fire. After saving your loved ones and pets, you have time to safely make a final dash to save any one item. What would it be? Why?
\end{enumerate} \\
\hline
\\
\centerline{\textbf{Ending}}
\vspace{-3pt}
\begin{enumerate}
\itemsep0em 
\item If you were able to live to the age of 90 and retain either the mind or body of a 30-year old for the last 60 years of your life, which would you choose?
\item If you could wake up tomorrow having gained one quality or ability, what would it be?
\end{enumerate}
\end{tabular}
\vspace{-8mm}
\end{table}
\vspace{-1.0mm}

\subsection{Crowdsourcing}
\vspace{-0.7em}
We posted our experiments as Human Intelligence Tasks (HITs) on \textit{Amazon Mechanical Turk}. HITs were defined as sets of utterances in either the contextual or random presentation condition.  In each condition, workers were presented with a single utterances and were asked to annotate the activation and valence values of that utterance using Self Assessment Manikins \cite{bradley1994measuring}. Once completed, the worker was presented with a new HIT and could not go back to revise a previous estimate of emotion. This annotation strategy is different than the one deployed in \cite{chen2018emotionlines},where the workers could go back and re-evaluate utterances. 


In the randomized experiment, each HIT is an utterance from any section, by any speaker, from any session and all HITs appear in random order. So, a worker might see the first HIT as \textit{Utterance 10 from Section 3 of Subject 4's stressed recording} and see the second HIT as \textit{Utterance 1 from Section 5 of Subject 10's non-stressed recording}. This setup ensured that the workers couldn't condition to any speaker's specific style or contextual information.

In the contextual experiment, we posted each HIT as a collection of ordered utterances from a section of a particular subject's recording. Because each section's question was designed to elicit a particular emotion, we still posted the HITs in a random order over sections from all subjects.  This prevented workers from conditioning to the speaking style of an individual participant. For example, a worker might see the first HIT as \textit{Utterance 1...N from Section 3 of Subject 4's stressed recording} and see the second HIT as \textit{Utterance 1...M from Section 5 of Subject 10's non-stressed recording} where {\it N, M} are the number of utterances in those sections respectively.

We recruited from a population of workers in the United States who are native English speakers, to reduce the impact of cultural variability.  We ensured that each worker had $>98$\% approval rating and number of HITs approved as $>500$. We ensured that all workers understood the meaning of activation and valence using a qualification task that asked workers to rank emotion content. The workers were asked to select, given two clips, which clip had the higher valence and which had the higher activation.  The options were chosen from a set including: (1) a speaker in low activation, high valence state and (2) a speaker in high activation, low valence state.  

We assigned each HIT to eight workers.  All HIT workers were paid a minimum wage ($\$9.25/$hr), pro-rated to the minute.  We removed and re-posted assignments where the worker completed the assignment in time shorter than the audio length.  The ground-truth for each utterance was formed by taking the average of the eight annotations.  
\vspace{0em}
\begin{table}[t]
  \caption{Data summary (R:random, C:context, F:female, M:male).}
  \footnotesize
  \label{data_stat}
  \vspace{1mm}
  \centering
  \begin{tabular}{ll}
    \toprule
    \multicolumn{2}{c}{\textbf{Monologue Subset}}\\
    \cmidrule(r){1-2}
    Mean num of utterances/monologue     & $9.69\pm2.55$    \\
    Mean duration of utterances       & $12.44 \pm 6.72$ seconds   \\
    Total num of utterances        & 2,648    \\
    Selected num of utterances     & 2,574    \\
    Gender distribution        & 19 (M) and 9 (F)   \\
    Total annotated speech duration & $\sim10$ hours\\
    \midrule
    \multicolumn{2}{c}{\textbf{Crowdsourced Data}}\\
    \cmidrule(r){1-2}
    Num of workers                 & 160 (R) and 72 (C)    \\
    Blocked Workers     & 8   \\
    \multirow{2}{*}{Mean activation}     & 3.62$\pm$0.91 (R)   \\
                                  &   3.69$\pm$0.81 (C)\\
    \multirow{2}{*}{Mean valence}       & 5.26$\pm$0.95 (R)   \\
     &    5.37$\pm$1.00 (C)\\
    \bottomrule
  \end{tabular}
  \vspace{-5mm}
\end{table}

    

\vspace{-4mm}

\section{Experimental Setup}
\label{exp_setup}
\vspace{-2mm}

\textbf{Acoustic Features.} We extract acoustic features using OpenSmile~\cite{eyben2010opensmile} with the eGeMAPS configuration~\cite{eyben2016geneva}. The eGeMAPS feature set consists of $88$ utterance-level statistics over the low-level descriptors of frequency, energy, spectral, and cepstral parameters. We perform speaker-level $z$-normalization on all features.\looseness=-1

\textbf{Static Network Setup (Hypothesis 4).} We train and evaluate four Deep Neural Networks (DNN) models: \{random, contextual\}$\times$ \{valence, activation\}. In all cases, we predict the continuous annotation using regression.  For each network setup, we follow a five-fold evaluation scheme and report the average RMSE across the folds. For each test-fold, we use the previous fold for hyper-parameter selection and early stopping. The hyper-parameters include: number of layers $\{2, 3, 4\}$ and layer width $\{64, 128, 256\}$. We use ReLU activation and train the networks with MSE loss using Adam optimizer. 

\textbf{Dynamic Network Setup (Hypothesis 5).} We use Gated Recurrent Unit networks (GRU). The hyper-parameters are: number of layers $\{1, 2\}$ and layer width $\{64, 128, 256\}$. We pass the GRU output of the last time step through a regression layer to get the final outputs. We train the networks with MSE loss using Adam optimizer.

\textbf{Network Training.} We train our networks for a maximum of 100 epochs and monitor the validation loss after each epoch. We stop the training if the validation loss does not improve for 15 consecutive epochs. We revert the network's weights to those that achieved the lowest validation loss during training. Finally we train each network five times and average the predictions to reduce variance due to random initialization.

\vspace{-1em}
\section{Results and Analysis}
\vspace{-0.7em}
\subsection{Question 1}
\vspace{-0.7em}
\uline{Hypothesis}: \textit{Human annotations collected through randomized labeling are significantly different from those collected through contextualized labeling.} Prior work has shown context effects emotion perception \cite{yannakakis2017ordinal}, even when observers are explicitly asked not to take it under consideration \cite{ngo2015use,cauldwell2000did}. Hence, we believe that context provided by previous utterances would lead to a change in perception of a particular utterance.
Tables~\ref{exp1-stress} and~\ref{exp1-section} (sets of significantly different means are bolded ($t$-test, $p<0.01$)) show the mean activation and valence, for the random and contextualized labeling schemes, grouped by condition and question, respectively. 
Table~\ref{exp1-stress} shows that, for non-stress conditions, the mean of the activation ratings obtained through contextual labeling is significantly higher than that obtained through random labeling. The table also shows that, for both stress and non-stress conditions, the valence means obtained through contextual labeling are significantly higher than those obtained through random labeling.
Table~\ref{exp1-section} shows that, although the mean valence and activation values were consistently different for the labelling schemes across all emotion elicitation techniques, the differences were significant in some elicitation techniques and not in others.


\begin{table}[t]
  \caption{Mean activation and valence values obtained from the two crowdsourcing labeling schemes (random and context) grouped by speaker condition (stress and non-stress).  
  }
  \label{exp1-stress}
  \vspace{1mm}
\footnotesize
  \centering
  \begin{tabular}{lcccc}
    \toprule
    & \multicolumn{2}{c}{Activation} & \multicolumn{2}{c}{Valence}  \\
    \cmidrule(r){2-3} \cmidrule(r){4-5}
         & Random     &  Context  & Random     &  Context\\
    \midrule
    Stress         & 3.63        & 3.59       & {\bf 5.27}  & \textbf{5.36}   \\
    Non-Stress     & {\bf 3.61}  & {\bf 3.79} & {\bf 5.26}  & {\bf 5.39}      \\
    \bottomrule
  \end{tabular}
  \vspace{-3mm}
\end{table}
\begin{table}[t]
  \caption{Mean activation and valence values obtained from the two crowdsourcing labeling schemes (random and context) grouped by emotion elicitation question.
  }
  \label{exp1-section}
  \vspace{1mm}
  \footnotesize
  \centering
  \begin{tabular}{lcccc}
    \toprule
    & \multicolumn{2}{c}{Activation} & \multicolumn{2}{c}{Valence}  \\
    \cmidrule(r){2-3} \cmidrule(r){4-5}
         & Random     &  Context  & Random     &  Context\\
    \midrule
    Icebreaker & 3.55  & 3.60 & {\bf 5.41}  & \textbf{5.61}    \\
    Positive   & 3.64  & 3.71 & 5.11  & 5.13      \\
    Negative   & {\bf 3.57}  & {\bf 3.67} & {\bf 5.40}  & {\bf 5.55}      \\
    Intensity  & {\bf 3.64}  & {\bf 3.74} & {\bf 5.17}  & {\bf 5.31}      \\
    Ending     & 3.69  & 3.71 & {\bf 5.23}  & {\bf 5.29}      \\
    \bottomrule
  \end{tabular}
  \vspace{-6mm}
\end{table}
\vspace{-3mm}
\vspace{-0.3em}
\subsection{Question 2}
\begin{figure}[t]
  \centering
  \caption{Mean difference between the self-reported activation and valence ratings and the random and contextual presentations.}
  \includegraphics[scale=0.50]{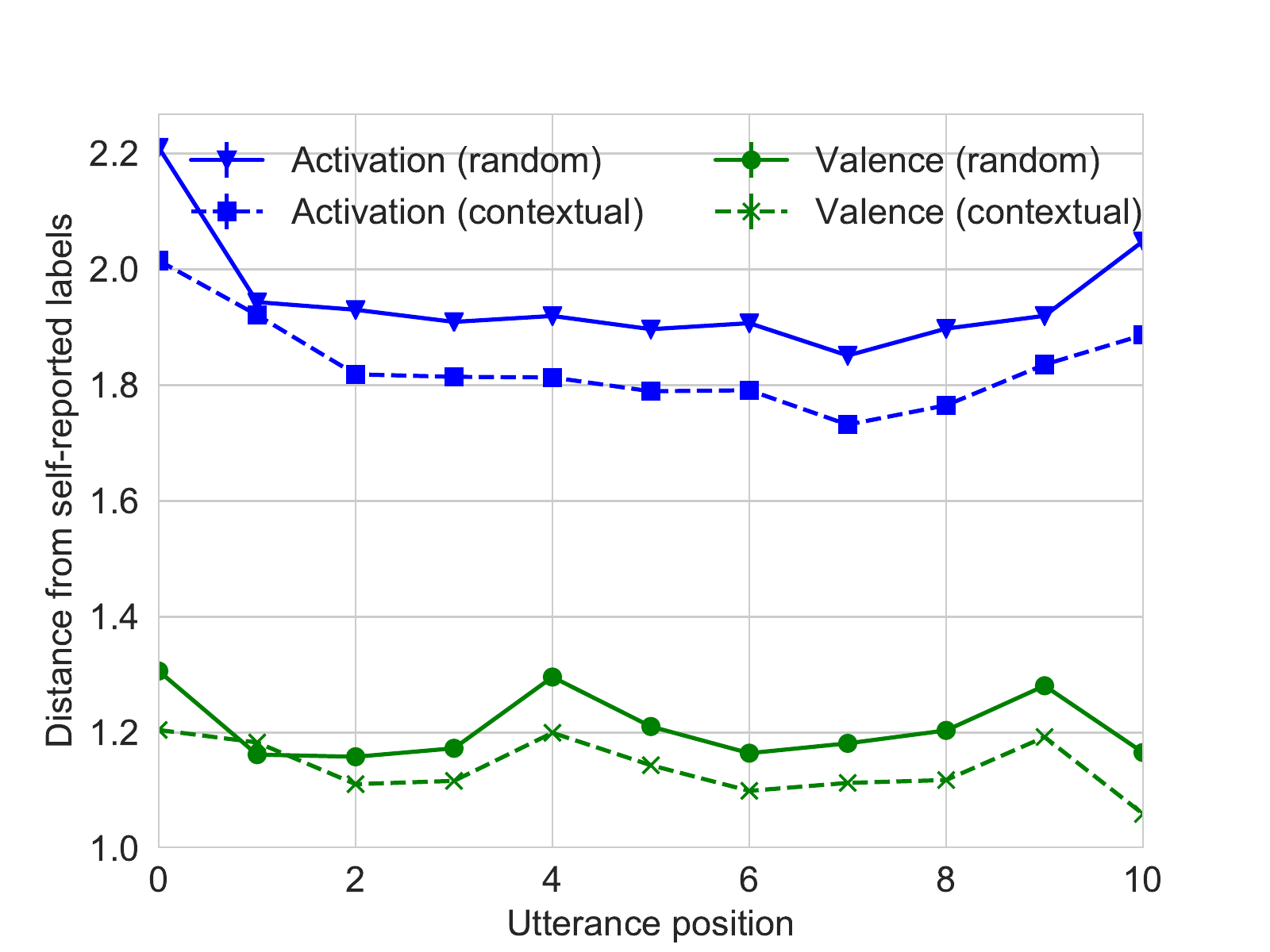}
  \vspace{-5mm}
  \label{fig:exp2}
\end{figure}

\vspace{-0.7em}
\uline{Hypothesis}: \textit{Annotations of outside observers are more similar to self-annotations in the contextual case, compared to the randomized case.} Path models \cite{banziger2015path} suggest that subjective voice variation, from the established mental baseline accounts for much of the variance in emotion inference. Hence, emotion inference is aided with more cues about the speech patterns that are more readily provided through context.
Figure~\ref{fig:exp2} shows the absolute differences between the mean crowdsourced labels (valence and activation, each for random and contextual schemes) and self-reported scores as a function of utterance position. The figure shows that contextual labels have consistently lower absolute differences, compared to self-reported labels, than the random labels. A paired $t$-test shows that these differences between the contextual and random labels are significant ($p<0.01$) for both valence and activation.

Our results suggest that crowdsourced emotion labels collected with access to contextual information are closer to self-reported emotion labels. Our results further suggest that these differences are consistent across recording conditions (Table~\ref{exp2-stress}) and emotion elicitation questions ( 
Table~\ref{exp2-section}, sets of significantly different means are bolded, $t$-test, $p<0.01$). 

\begin{table}[t]
    \footnotesize
    \caption{Mean difference between the self-reported activation and valence ratings from the two labeling schemes (random and context) grouped by speaker condition (stress and non-stress).
    }  
  \label{exp2-stress}
  \vspace{1mm}
  
  \centering
  \begin{tabular}{lcccc}
    \toprule
    & \multicolumn{2}{c}{Activation} & \multicolumn{2}{c}{Valence}  \\
    \cmidrule(r){2-3} \cmidrule(r){4-5}
         & Random     &  Context  & Random     &  Context\\
    \midrule
    Stress         & {\bf 2.03}        & {\bf 1.96}       & {\bf 1.20}  & \textbf{1.14}   \\
    Non-Stress     & {\bf 1.82}  & {\bf 1.67} & {\bf 1.20}  & {\bf 1.12}      \\
    \bottomrule
  \end{tabular}
  \vspace{-1.8em}
\end{table}

\begin{table}[t]
  \caption{Mean difference between the self-reported activation and valence ratings from the two labeling schemes (random and context) grouped by emotion elicitation question.  
  }
  \label{exp2-section}
  \footnotesize
  \vspace{1mm}
  \centering
  \begin{tabular}{lcccc}
    \toprule
    & \multicolumn{2}{c}{Activation} & \multicolumn{2}{c}{Valence}  \\
    \cmidrule(r){2-3} \cmidrule(r){4-5}
         & Random     &  Context  & Random     &  Context\\
    \midrule
    Icebreaker & 1.81  & 1.80 & {\bf 0.97}  & \textbf{0.85}    \\
    Positive   & {\bf 1.89}  & {\bf 1.74} & 1.14  & 1.11      \\
    Negative   & {\bf 1.96}  & {\bf 1.76} & {\bf 1.18}  & {\bf 1.07}      \\
    Intensity  & {\bf 2.19}  & {\bf 2.08} & {\bf 1.49}  & {\bf 1.44}      \\
    Ending     & {\bf 1.81}  & {\bf 1.73} & 1.23  & 1.28      \\
    \bottomrule
  \end{tabular}
  \vspace{-2em}
\end{table}



\vspace{-1.3em}
\subsection{Question 3}
\vspace{-0.7em}
\uline{Hypothesis}: \textit{Individual annotators differ in annotation similarity in the contextual presentations, compared to the randomized presentation.} Joseph et al. in \cite{joseph2017constance} show that while insufficient context results in noisy and uncertain annotations, an overabundance of context may cause the context to outweigh other signals and lead to lower agreement. Further, contextual information biases different people differently on both temporal and intensity metrics \cite{van2009immediacy,walker2009fading}.  Our results highlight the impact of context: the agreement is significantly higher in the case of labels obtained from the randomized presentations, compared to the contextualized presentations: ($1.55$ vs. $1.62$) for activation and ($1.07$ vs. $1.14$) for valence. This trend holds true for all experimental design setups i.e. \{random, contextual\}$\times$ \{valence, activation\} and \{random, contextual\}$\times$ \{icebreaker, positive, negative, intensity and ending\}. As shown in Tables~\ref{exp1-stress} and \ref{exp1-section}, the labels obtained in both cases are significantly different due to context-based conditioning. However, the conditioning may not impact the labels consistently across all workers, which may lead to lower inter-annotator agreement values. This suggests that it may be beneficial to consider the distribution of annotations as ground-truth, rather than averaging labels, which presumes that the impact of conditioning is consistent across all workers \cite{zhang2017predicting}.

\vspace{-4mm}
\subsection{Question 4}
\vspace{-0.7em}
\uline{Hypothesis}: \textit{A static classifier will perform better when trained and evaluated using labels annotated with a randomized presentation, compared to a  contextualized presentation.} Prior studies have shown that it is easier to classify data with less target variation \cite{liu2016classification} and matched classifier input, which in our case is labels obtained from the random labelling presentation (the classifier processes single utterances at a time, no context).

We test this hypothesis by training and evaluating classifiers for the four possible setups: $\{random,$ $ contextual\}$ x $\{valence, $ $activation\}$. The classifier is described in Section~\ref{exp_setup}.  We find that the RMSEs are lower for the contextual labels in the case of activation ($0.91$ vs. $1.00$) while the errors are lower for the random labels in the case of valence ($1.13$ vs. $1.20$). Using a paired $t$-test, we find that the differences in errors are significant in the case of valence but not activation. 
These findings suggest that classification performance is impacted by the labelling methodology, but that this effect may depend on emotion dimension.  

Prior work has demonstrated the importance of considering long-term context when predicting valence (the same effect has not been shown in activation)~\cite{khorram2017capturing}.  The contextual annotations provided the annotators with this information, but the classifier could not take advantage of this effect.  This mismatch may have contributed to the relatively lowered performance of the valence classifier, compared to the activation classifier.


\vspace{-3.5mm}
\subsection{Question 5}
\label{q5}
\vspace{-0.7em}
\uline{Hypothesis}: \textit{We anticipate that systems trained on contextualized labels will see greater increases in performance as the amount of provided context increases.}  This finding would support results in the literature regarding the ordinal nature of emotion perception \cite{yannakakis2017ordinal} and previous works in emotion recognition that have demonstrated that context can influence the performance of emotion classifiers~\cite{khorram2017capturing}. 

The classifier is described in Section~\ref{exp_setup}.  We test this hypothesis by using the contextual annotations in one classifier and the non-contextual (random) annotations for the other classifier.
We select a subset of utterances in each section that have at least five consecutive utterances before them ($59\%$ of the original data). The initial classifier is trained without temporal context (but with the contextualized labels).  We incrementally increase the number of past utterances (from zero to five). We run this for every task combination and report the results in Table~\ref{exp5-errors}.

Table~\ref{exp5-errors} shows the performance gains after incrementally adding the past utterance, relative to the baseline performance. The addition of past utterances improves the performance over baseline for all setups. Where using contextual labels, however, the performance gains are generally higher than the gains obtained after using random labels. Our results suggest that it is necessary to consider the mismatch the amount of context provided to the annotators generating the labels and the ultimate classification system.
\vspace{-0.9em}

\begin{table}[t]
  \caption{Relative improvement in RMSE (\%) obtained for each additional previous utterance, comparing random and contextual labels.}
  \label{exp5-errors}
  \vspace{1mm}
  \footnotesize
  \centering
  \begin{tabular}{ccccc}
    \toprule
     & \multicolumn{2}{c}{Activation} & \multicolumn{2}{c}{Valence}  \\
    \cmidrule(r){2-3} \cmidrule(r){4-5}
         Past steps & Random     &  Context  & Random     &  Context\\
    \midrule
    0 & -    & -    & -    & -\\
    1 & $+1.96\%$ & $+1.24\%$ & $+0.85\%$ & $+3.32\%$\\
    2 & $+2.28\%$ & $+2.93\%$ & $+5.23\%$ & $+7.63\%$\\
    3 & $+3.36\%$ & $+8.72\%$ & $+6.08\%$ & $+8.43\%$\\
    4 & $+4.41\%$ & $+10.5\%$ & $+8.23\%$ & $+8.36\%$\\
    \bottomrule
  \end{tabular}
  \vspace{-2em}
\end{table}
\section{Conclusion}
\vspace{-0.9em}
In this work we showed that the amount of context provided to annotators when assigning emotion labels affects both the annotations themselves and the performance of classifiers using these annotations. We also studied the implications of a mismatch between annotation context and classifier context on classifier performance. For future work, we will analyze the effect of context given multimodal information and the differences in perception of emotion expression in stress vs. non-stressed situations.

\vspace{-0.2em}
\begin{center}
    \textbf{ACKNOWLEDGEMENTS}
\end{center}
\vspace{-0.5em}

This material is based in part upon work supported by the Toyota Research Institute (``TRI''), the IBM PhD Fellowship Award, and by the National Science Foundation (NSF CAREER 1651740). Any opinions, findings, and conclusions or recommendations expressed in this material are those of the authors and do not necessarily reflect the views of the NSF, TRI, or any other Toyota entity. The authors would also like to thank Veronica Perez for her input and help in collecting the dataset. \looseness -1
\vspace{-5mm}
\bibliographystyle{IEEEbib}
\bibliography{main} 
\looseness=-1

\end{document}